\documentclass[research]{fcs}

\usepackage{bm}
\usepackage{times}
\usepackage{soul}
\usepackage{url}
\usepackage[utf8]{inputenc}
\usepackage{subfigure}
\usepackage{graphicx}
\usepackage{amsmath}
\usepackage{amsthm}
\usepackage{amssymb}
\usepackage{booktabs}
\usepackage{algorithmic}
\usepackage[ruled,linesnumbered,noend]{algorithm2e}

\usepackage{multirow}
\usepackage[normalem]{ulem}
\useunder{\uline}{\ul}{}
\usepackage{enumitem}
\usepackage{enumerate}
\usepackage{color}
\usepackage{microtype}

\newcommand\norm[1]{\left\lVert#1\right\rVert}
\newcommand{\Lapl}{\mathbf{\mathop{\mathcal{L}}}}

\newcommand{\Trans}[1]{{#1}^{\top}}

\newcommand{\Mat}[1]{\mathbf{#1}}

\newcommand{\Space}[1]{\mathbb{#1}}
\newcommand{\Set}[1]{\mathcal{#1}}

\newcommand{\ie}{\emph{i.e., }}
\newcommand{\eg}{\emph{e.g., }}

\title{Rumor Detection with Self-supervised Learning on Texts and Social Graph}
\author[1]{Yuan Gao}
\author*[1]{Xiang Wang}
\author*[1]{Xiangnan He}
\author[2]{Huamin Feng}
\author[1]{Yongdong Zhang}
\address[1]{School of Information Science and Technology, University of Science and Technology of China, Hefei 230026, China}
\address[2]{Beijing Electronic Science and Technology Institute, Beijing 102627, China}
\fcssetup{
  received       = {month dd, yyyy},
  accepted       = {month dd, yyyy},
  corr-email     = {xiangnanhe@gmail.com,xiangwang1223@gmail.com},
}
\begin{abstract}
    Rumor detection has become an emerging and active research field in recent years.
    At the core is to model the rumor characteristics inherent in rich information, such as propagation patterns in social network and semantic patterns in post content, and differentiate them from the truth.
    However, existing works on rumor detection fall short in modeling heterogeneous information, either using one single information source only (\eg social network, or post content) or ignoring the relations among multiple sources (\eg fusing social and content features via simple concatenation). \\
    Therefore, they possibly have drawbacks in comprehensively understanding the rumors, and detecting them accurately.
    In this work, we explore contrastive self-supervised learning on heterogeneous information sources, so as to reveal their relations and characterize rumors better.
    Technically, we supplement the main supervised task of detection with an auxiliary self-supervised task, which enriches post representations via post self-discrimination.\\
    Specifically, given two heterogeneous views of a post (\ie representations encoding social patterns and semantic patterns), the discrimination is done by maximizing the mutual information between different views of the same post compared to that of other posts.
    We devise cluster-wise and instance-wise approaches to generate the views and conduct the discrimination, considering different relations of information sources.
    We term this framework as \emph{Self-supervised Rumor Detection} (SRD).
    Extensive experiments on three real-world datasets validate the effectiveness of SRD for automatic rumor detection on social media.
\end{abstract}

\begin{document}
\section{Introduction}
Social media platforms (\eg Twitter, Facebook, Sina Weibo) have emerged as a major source of information, where users can easily view, post, forward, and comment on any content at little price.
However, it also causes the proliferation of rumors, which encode misinformation and disinformation.
The massive spread of rumor circulation has become a serious threat, influencing the societal or political opinions of the public.
Taking the 2016 US presidential election as an example, the inflammatory news that favored two nominees was shared over 37 million times on Facebook \cite{Farajtabar2017}, while around 19 million malicious bot accounts tweeted or retweeted supporting two nominees on Twitter \cite{DBLP:conf/sbp-brims/JinCGZWL17}.
Therefore, it is of great importance to detect rumors accurately and debunk rumors early before they reach a broad audience.

Identifying and modeling the patterns of rumor posts has become the theme, so as to differentiate the rumors from the truths.
Towards this end, multiple information sources are leveraged to profile the rumors and truths.
Among them, two informative and useful sources are (1) social network, which consists of user-user social connections and user-post interaction connections, reflecting how information flows and propagates along with these connections; and (2) post content, which are typically composed of text pieces and often associated with pictures or videos, that encode rich semantics.
Here we focus mainly on these semantic and propagation patterns.

Existing works on rumor detection follow a supervised learning scheme --- generating post representations from these information sources and feeding them into a supervised learning model guided by the ground-truth labels.
Learning post representations roughly falls into two types:
(1) using one information source individually;
(2) fusing two information sources.
Nonetheless, both types of methods suffer from some inherent limitations:
\begin{itemize}
    \item There are numerous efforts on modeling either post content or social network solely, but failing to model them simultaneously.
    Specifically, a line considers the semantic patterns (\eg lexical, syntactic, sentiment, topic-level features) in post content \cite{rucha2017csi,shu2019defend}, while another line focuses on the propagation patterns in social network \cite{Bian2020,Wu2020rumor}, in order to generate high-quality representations of rumors.
    For example, earlier work \cite{Ma2018} proposes to formulate the rumor propagation as a sequence via a recursive neural network. More recently, Bi-GCN \cite{Bian2020} employs graph neural network (GNN) over the holistic social network to perform the information propagation and aggregation among relevant posts.
    While being effective from the perspective of modeling individual information source, these methods leave other supplementary information untouched, which degrade model effectiveness.
    \item Some works \cite{ijcai2020-197,Nguyen_2020} explore how to integrate multiple information sources to characterize rumors more comprehensively. For example, CGAT \cite{ijcai2020-197} combines the representations encoding propagation and semantic patterns together via concatenation and pooling operators.
    Nevertheless, such integration is of limited ability to model the relations between these co-occurring heterogeneous patterns. It has been proved that information with different contents has different audiences and different propagation paths which makes different temporal patterns and diffusion speeds\cite{Ahmad2019rumor}.
    Taking the 2016 US presidential election as an example, given the content of a post supporting one nominee, we could possibly infer the audiences of interest and their sentiments, and further imagine its propagation among these audiences; analogously, we could estimate the coarse-grained topic of a post, based on its information flow.
    Hence, ignoring the relations falls short in revealing hidden patterns and results in suboptimal representations of rumors.
    
\end{itemize}

\begin{figure}[!t]
    \centering
    \includegraphics[width=\columnwidth]{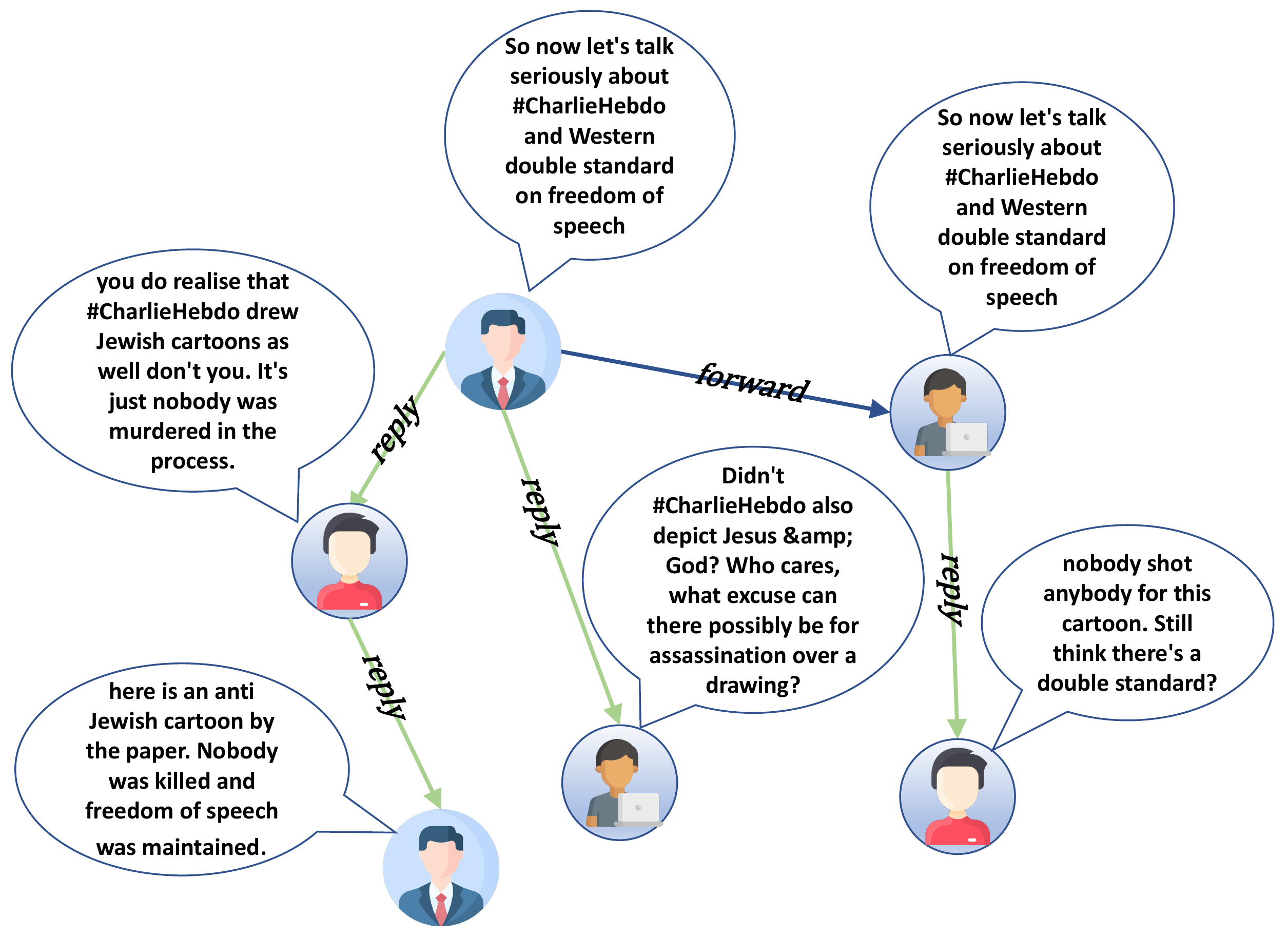}
    \label{fig_pheme}
    \vspace{-10pt}
    \caption{Real rumor in PHEME dataset\cite{Zubiaga2016}, which is expressed as propagation tree with text content.}
    \vspace{-10pt}
\end{figure}

To overcome the limitations of prior studies, we explore self-supervised learning (SSL) \cite{Ting2020,devlin-etal-2019-bert,he2020momentum} on heterogeneous information sources, to reveal their relations and characterize rumors better.
Technically, we supplement the main supervised learning task of rumor detection with an auxiliary self-supervised learning task of post self-discrimination.
These two tasks are founded upon the same post representations --- we apply the GNN encoder over the social network and aggregate the information from the neighbors, so as to inject the propagation patterns into the social representation of a post; meanwhile, we employ the CNN encoder to generate the semantic representations from a post's content.
The SSL task is composed of two modules:
(1) Data augmentation: we devise instance-wise and cluster-wise augmentation operators, and leverage them to derive different views of a post from its social and semantic representations;
(2) Contrastive learning: we encourage the agreement between different views of the same post to be maximized, while minimizing the agreement between different posts.
Such an SSL task allows us to explicitly measure the relations among different information sources, and guides the learning of post representations.
We term this framework as \emph{Self-Supervised Rumor Detection} (SRD).
Extensive experiments are conducted on three real-world datasets.
Experimental results from Twitter and Weibo demonstrate superior performance over the state-of-the-art methods like Bi-GCN \cite{Bian2020}.
Our contribution can be summarized as follows:
\begin{itemize}
\item We integrate the propagation and semantic patterns of posts via self-supervised learning. To the best of our knowledge, we are the first to leverage self-supervised learning for rumor detection on social media.
\item We propose cluster-wise and instance-wise discrimination as the self-supervised learning task to fulfill the potential of heterogeneous data.
\item Extensive experiments on three real-world datasets demonstrate the superiority of our proposed framework. We will release our codes upon acceptance.
\end{itemize}

\section{Related Work}
In this section, we review the work on rumor detection, graph convolutional network, and self-supervised learning.

\subsection{Rumor Detection}
Recently, automatic rumor detection comes into public attention. Traditional methods mainly focus on machine learning algorithms such as decision trees and SVM, which utilize manually extracted discrete features to determine if the sample given is worth trusting. Castillo et al.\cite{Castillo2011} identifies four kinds of features from Twitter, message, user, topic, and propagation, respectively. Following-up works mainly focus on characterizing one or a mixture of the 4 types of features. Yang et al.\cite{Yang2012} studies two new features, the client program used and event location, and extends the original work to Chinese. Ma et al.\cite{Ma2015} takes time stamps for news events into account and designs a dynamic series-time structure to model the changes or the trends of individual messages along the life cycle. Liu et al.\cite{Liu2015} points out the importance of real-time prediction, and found some features are more important for early detection while others begin dominating in later stages. Ma et al.\cite{Ma2017} adopts propagation trees to represent the propagation of each source tweet, and captures sub-structures that are indicative of rumors by estimating the similarity between propagation trees via propagation tree kernel learning. The performance of these models largely relies on the quality of feature engineering.

With the rise of deep learning and neural network, the focus has shifted from feature engineering to model learning. Ma et al.\cite{Ma2016} employs Recurrent Neural Network (RNN) to model news event as a sequence to better capture contextual information. Jin et al.\cite{Jin2017} fuses features from textual, visual, and social context contents, and utilizes attention mechanism to align features in RNN. Chen et al. \cite{Chen2018} embeds attention mechanism to RNN to enable importance focus that varies over time. Some researchers explore adding additional features to the model. Guo et al. \cite{guo2018rumor} adopt some social features and design a hierarchical social attention network. And Li et al. \cite{li2019rumor} add user profiles to evaluate the user credibility. Ma et al.\cite{Ma2018} proposes tree-structured Recursive Neural Networks (RvNN) to make predictions in a hierarchical structure with features extracted from both text contents and propagation structures. Yu et al.\cite{Yu2017} adopts a Convolutional Neural Networks (CNN) based model to automatically obtain key features from the representation learned from an unsupervised method - paragraph vector. Kumar et al.\cite{Kumar2020} introduces both a CNN and a bidirectional LSTM ensembled network with an attention mechanism to do rumor detection. Similarly, Liu et al\cite{Liu2018} detects fake news using a combination of recurrent and convolutional networks. Rao et al\cite{rao2021stanker} pre-trains a two-layer BERT, which takes comments as auxiliary features for fine-tuning. Recently, adversarial learning is proved to be helpful with the robustness of the model. Song et al\cite{song2021adversary} devises adversarial learning to improve the vulnerability of detection models. DropAttack\cite{li2021meet} is a new adversarial training method, which allows a certain weight parameter of the model to be attacked with a certain probability. These methods make much progress in misinformation identification, but they neglect the global structural features of rumor propagation. 

\subsection{Graph Convolutional Network}
Inspired by the great success of Convolutional Neural Network (CNN), Graph Neural Networks (GNNs) begin to emerge in supervised or semi-supervised tasks like node classification\cite{Kipf2017}, link prediction\cite{zhang2018link} and graph classification\cite{Pan2016}, as well as unsupervised tasks like network embedding\cite{wang2016kdd} and graph generation\cite{you2018graphrnn}. 

Graph Neural Network can be categorized into two directions: the spectral-based approaches and the spatial-based approaches\cite{Wu_2021}. The spectral-based approaches are based on spectral graph theory, they inherit the idea from graph signal processing and propose variant graph convolution kernels. The spectral-based methods are powerful but most of them are in transductive settings. The spatial-based methods construct convolutional filters by looking into the spatial relationship between nodes. Since their focus is on training a function that aggregates neighborhood information to get the representation of the current node, spatial-based methods generalize GNN to inductive learning. 

Kipf et al.\cite{Kipf2017} proposes a fast approximate convolution on graphs, which becomes one of the most representative works in Spectral GCN. Niepert et al.\cite{pmlr-v48-niepert16} discusses the equivalence of neighborhood graph and receptive field. Hamilton et al. \cite{hamilton2018inductive} proposes GraphSAGE, an inductive framework that is able to deal with previously unseen data. To address the issue that neighborhood contribution may not be the same, veličković et al.\cite{veli2018graph} proposes Graph Attention Network (GATs) that introduces attention mechanism to neighborhood aggregation. Due to their advantages in dealing with non-euclidean data, GNNs have proved their power in areas like recommender systems\cite{he2020lightgcn,Wang_2019,zeng2022shadewatcher}, social network\cite{hamilton2018inductive} that always make contact with graph-structured data. In rumor detection, Tian et al.\cite{Bian2020} proposes Bi-GCN, with top-down direction for rumor propagation and bottom-up direction for rumor dispersion; Yang et al.\cite{ijcai2020-197} constructs four types of camouflage strategies which are introduced to GCN through adversarial training; Lin et al. \cite{lin2021rumor} designs a Hierarchical Graph Attention Networks to attend over the responsive posts that can semantically infer the target claim.

\subsection{Self-supervised Learning}
Deep neural network models can solve supervised learning problems well with enough labeled data. However, manual labels are expensive and may require domain-specific knowledge. Self-supervised learning can generate pseudo-labels, or cluster input through the comparison between positive and negative pairs. Thus studies on self-supervised learning are divided into two branches: \textit{generative models}\cite{devlin-etal-2019-bert}\cite{NIPS2016_b1301141} and \textit{contrastive models}\cite{hjelm2019learning}.

For generative model variants, one of the most popular models is auto-encoding. This method encodes the input data, and tries to reconstruct it with decoding, during the process noises can be intentionally added to enhance model robustness\cite{devlin-etal-2019-bert}. Ian et al\cite{goodfellow2014generative} proposes a generative adversarial network (GAN) for estimating generative models via an adversarial process.

Contrastive models are another important branch in self-supervised learning. Michael et al.\cite{pmlr-v9-gutmann10a} defines the notion of Noise Contrastive Estimation (NCE), acting as the objective of most contrastive learning models. The comparison objects can be either context-instance contrast or instance-instance contrast\cite{liu2021selfsupervised}. The former models the relationship between the local part of a sample and its global context representation. Hjelm et al.\cite{hjelm2019learning} proposes Deep Infomax (DIM) that maximizes the mutual information between the local part of an image and its global context. Contrastive Predictive Coding (CPC)\cite{oord2019representation} generalizes this work to speech recognition, which contrasts audio segments with the whole audio. However, Tschannen et al.\cite{tschannen2020mutual} provides empirical evidence that MI may not be the only reason for this success, and it also strongly depends on the encoding structure and negative sampling strategy. In this case, instance-instance contrastive focus on instance-level representation instead of MI. One direction\cite{caron2019deep}\cite{caron2021unsupervised}\cite{Humam2020} iterates between feature encoding and cluster discrimination. The cluster assignment from the discriminator will act as pseudo label signal to improve the quality of the encoder. Another branch is instance-discrimination. He et al. \cite{he2020momentum} proposes Momentum Contrast (Moco) which substantially increases the number of negative samples. Chen et al.\cite{Ting2020} proposes SimCLR that proves the importance of hard positive example, the model follows the end-to-end framework and is trained in quite large batch size. Our method mainly focuses on contrastive learning.

There are also some recent works on SSL for graph data. Veličković et al.\cite{veli2018deep} Deep Graph Infomax (DGI), which extends DIM to graph-structured data. Kim and Oh\cite{kim2021how} proposes superGAT which self-supervise graph attention through edges. Wu et al.\cite{wu2020selfsupervised} applies graph SSL to recommender systems, and presents theoretical analyses of \\ self-supervised graph learning. However, to the best of our knowledge, self-supervised learning has not been explored in the task of rumor detection.
\section{Methodology}

In this section, we first formulate rumor detection problems (Section 3.1), and introduce 4 core parts of our proposed framework SRD (Section 3.2).

\subsection{Task Description}
We first introduce some commonly used notations. We would use bold capital letters (e.g. \textbf{\emph{A}}) to represent matrices and lowercase letters (e.g. \textbf{\emph{v}}) to represent vectors. Note that by default all of the vectors are in column, i.e. $\textbf{\emph{v}}\in \mathbb{R}^{d}$. $\emph{X}_{ij}$ stands for the element in matrix $\textbf{\emph{X}}$ located at row $i$ and column $j$. Table \ref{table:0} lists some of the terms mentioned.

As an individual post contains very limited context and is short in nature, we follow previous works\cite{Bian2020,Ma2016} and focus on rumor detection at a granular level of event (\ie a group of posts) instead.
Let $\Set{C}$ be a set of events, where each event $c\in\Set{C}$ originates from a source post with the rumor or truth label $y_{c}$.
Event $c$ is associated with two heterogeneous data:
(1) propagation tree $<\Set{V},\Set{E}>$, where $\Set{V}$ and $\Set{E}$ denote the sets of post nodes and their connections.
In particular, the source post of $c$ serves as the root node, and the other nodes are the replies relevant to the source post.
The edge $(v_{i},v_{j})\in\Set{E}$ is directed, representing that post $v_{i}$ is a reply to $v_{j}$.
(2) semantic contents, where each post node $v\in\Set{V}$ is attached with its content features $\Mat{x}\in\Space{R}^{d}$, and $d$ is the feature dimension.
As a result, we can reorganize each event $c$ in the form of an attributed graph with the adjacency matrix $\Mat{A}\in\Space{R}^{|\Set{V}|\times|\Set{V}|}$, whose element is:
\begin{gather}
    A_{ij}=
    \begin{cases}
     1,\quad (v_{i},v_{j})\in\Set{E}\\
     0,\quad \text{otherwise}
    \end{cases}.
\end{gather}
We formulate the task of rumor detection as follows:
given an event $c$ with an attributed graph, we would like to exploit its heterogeneous data (\ie propagation tree and semantic content), and predict whether the information contained in the event is credible.

\subsection{Our Framework}
In this section, we present our proposed framework, self-supervised rumor detection (SRD).
It is composed of four modules: (1) propagation representation learning, which applies a GNN model on the propagation tree; (2) semantic representation learning, which employs a text CNN model on the post contents; (3) contrastive learning, which models the co-occurring relations among propagation and semantic representations; and (4) rumor prediction, which builds a predictor model upon the event representations.

\begin{table}[!t]
\caption{Notations and Corresponding Explanations}
\label{table:0}
\centering
\resizebox{\columnwidth}{!}{
\begin{tabular}{c|l}
 \hline
 Notation & Explanation \\ 
 \hline\hline
 $\textbf{\emph{A}}, \textbf{\emph{D}}$ & Adjacency matrix and degree matrix of a graph\\

 $\textbf{\emph{X}}=[\textbf{\emph{x}}_{1}, \textbf{\emph{x}}_{2}, \cdots , \textbf{\emph{x}}_{N}]^{T}$ & Feature matrix of the attributed graph\\

 $\textbf{\emph{W}}^{l}, \textbf{\emph{b}}^{l}$ & Trainable weights and bias term at layer $l$\\

 $\theta$ & Model parameters requires gradient\\
 
 $g_{i}$ & graph view representation\\
 $t_{i}$ & text view representation\\\hline
\end{tabular}}
\end{table}

\begin{figure*}[!t]
\centering
\includegraphics[width=\textwidth]{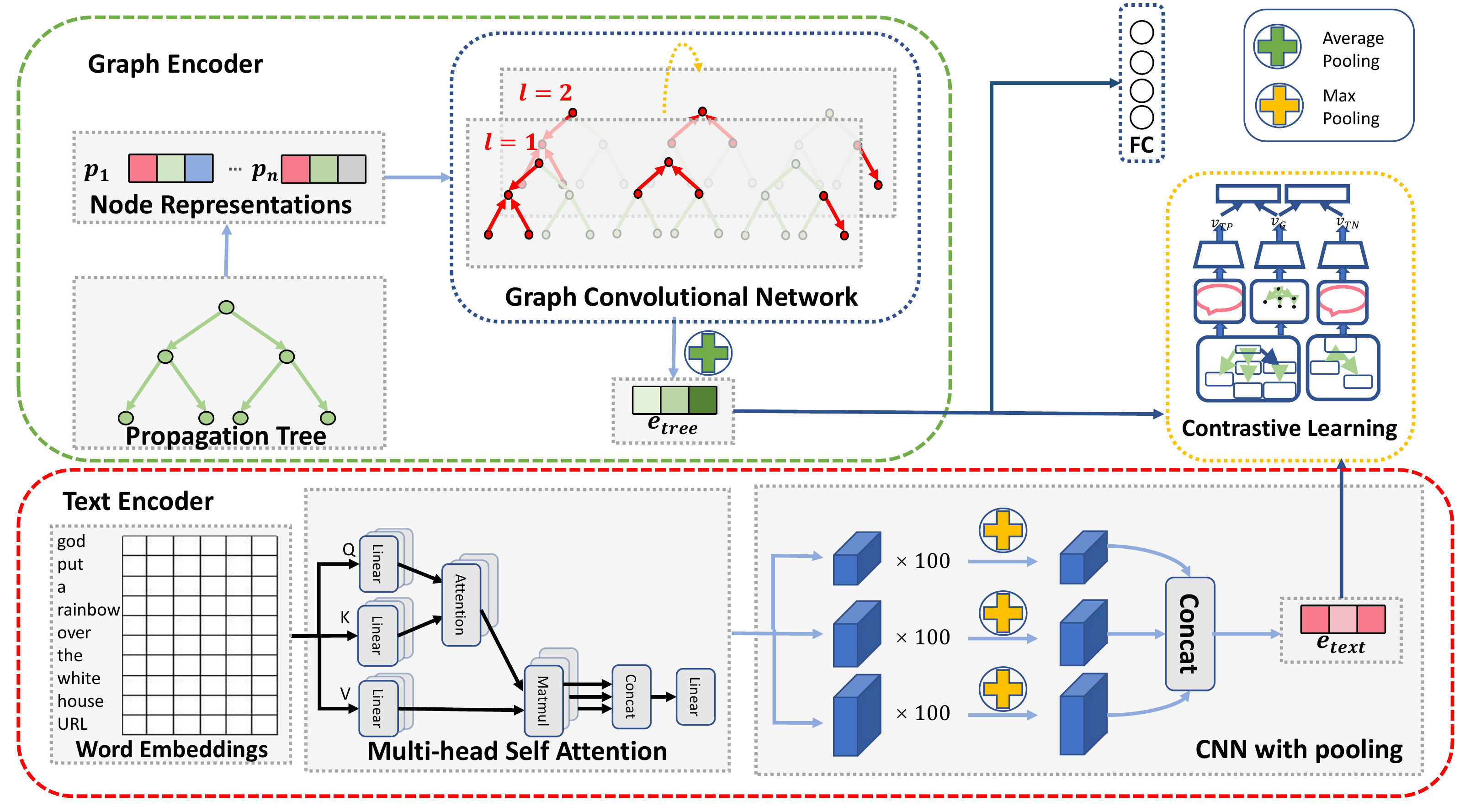}
\caption{The framework of our proposed self-supervised learning rumor detection (SRD). The framework consists of 4 modules: Graph Encoder, Text Encoder, Contrastive Learning Module, and FC supervised prediction module. The contrastive learning module here is thumbnail of Fig \ref{fig_dc}.}
\label{fig_topo}
\end{figure*}

\subsubsection{Propagation Representation Learning}
The propagation tree of a post presents how its information flows along with the source network, especially reflecting the user-user interactions.
Prior studies \cite{yang2015detecting} have shown that the propagation patterns of rumors are different from that of the truth.
Inspired by this, we use a graph neural network (GNN) to extract such patterns as propagation representations of posts.
At its core is to apply the neighborhood aggregation scheme on the propagation tree $\Mat{A}$ of a post, which updates the node representations by aggregating the vectorized information passing from neighbors:
\begin{gather}\label{equ:gnn-layer}
    \Mat{H}^{(l)} = \sigma(\Mat{D}^{-\frac{1}{2}}\hat{\Mat{A}}\Mat{D}^{-\frac{1}{2}}\Mat{H}^{(l-1)}\Mat{W}^{(l)}),
\end{gather}
where $\Mat{H}^{(l)}$ is the node representation at the $l$-th layer, $\Mat{H}^{(l-1)}$ is that of previous layer, and $\Mat{H}^{(0)}=\Mat{X}$ uses the multi-hot encoding embedding in terms of word distribution in post as the initial representations;
$\Mat{D}$ is the degree matrix, and $\hat{\Mat{A}}=\Mat{A}+\Mat{I}$ is the adjacency matrix with the identify matrix $\Mat{I}$ representing self-connections;
$\sigma(\cdot)$ is the nonlinear activation function, which we set as ReLU.
After adopting $L$ layers recursively, we finalize the representations as $\Mat{H}^{(L)}$, which encode $L$-order connectivity among nodes.
Here we set $L$ as $2$.

Having established the representations of nodes within the propagation tree, we use a readout function to get the representation of the source post.
Here we resort to the mean pooling operator $f_{\text{mean-pooling}}(\cdot)$:
\begin{gather}\label{equ:gnn-pooling}
    \Mat{g}=f_{\text{mean-pooling}}(\Mat{H}^{(L)}).
\end{gather}

\subsubsection{Semantic Representation Learning}
Most of the time in human society, authorities and experts identify rumors by looking into the textual materials to find suspicious content based on their domain knowledge. Intuitively, semantic features have implications for the credibility of a post.
To address this issue, we adopt a multi-head self-attention layer combined with CNN to get a comprehensive representation of the post\cite{ijcai2020-197}. Here powerful pre-train models like BERT are not chosen for two reasons: (1) The size of the benchmark datasets are too small to do fine-tuning. (2) They are time-consuming and need a great amount of memory. 

Multi-head attention\cite{NIPS2017_3f5ee243} allows for capturing word representations from different subspaces.
Considering a multi-head attention layer with $h$ heads, each head is a parallel attention layer.
These heads share three input matrices: query matrix $\Mat{Q}$, key matrix $\Mat{K}$, and value matrix $\Mat{V}$.
For each head $i$, these input matrices will be projected into $d_{k}$, $d_{k}$, $d_{v}$ dimension subspaces as $\textbf{\emph{Q}}_{i}$, $\textbf{\emph{K}}_{i}$, $\textbf{\emph{V}}_{i}$ through trainable linear projections $\textbf{\emph{W}}_{i}^{Q} \in \mathbb{R}^{d_{model}\times d_{k}}$, $\textbf{\emph{W}}_{i}^{K} \in \mathbb{R}^{d_{model}\times d_{k}}$, $\textbf{\emph{W}}_{i}^{V} \in \mathbb{R}^{d_{model}\times d_{v}}$, respectively. Here we adopt the self-attention scheme where $\textbf{\emph{Q}}_{i} = \textbf{\emph{K}}_{i} = \textbf{\emph{V}}_{i}$.
This scheme enables every word to attend to each other, and as a result, every single word will be represented by the most similar words.
More formally, at each iteration, vectors in $\textbf{\emph{Q}}_{i}$ serve as \textit{queries} and interact with \textit{keys} in $\textbf{\emph{K}}_{i}$ via inner product scaled by $\sqrt{d_{k}}$, on which a softmax function is applied to obtain weights for \textit{values} in $\textbf{\emph{V}}_{i}$. We denote the output of the $i$-th head by $\textbf{\emph{Z}}_{i}$:
\begin{equation}\label{eq:4}
    \textbf{\emph{Z}}_{i} = f_{\text{attention}}(\textbf{\emph{Q}}_{i}, \textbf{\emph{K}}_{i}, \textbf{\emph{V}}_{i}) = f_{\text{softmax}}(\frac{\textbf{\emph{Q}}_{i}\textbf{\emph{K}}_{i}^{T}}{\sqrt{d_{k}}})\textbf{\emph{V}}_{i}.
\end{equation}

The final output \textbf{\emph{Z}} of multi-head attention can be calculated as a linear projection of concatenation of \textit{h} heads:
\begin{equation}\label{eq:mhout}
    \textbf{\emph{Z}} = f_{\text{multi-head}}(\textbf{\emph{Q}}, \textbf{\emph{K}}, \textbf{\emph{V}}) = f_{\text{concatenate}}(\textbf{\emph{Z}}_{1}, ... , \textbf{\emph{Z}}_{h})\textbf{\emph{W}}^{O},
\end{equation}
where $\textbf{\emph{W}}^{O} \in \mathbb{R}^{hd_{v}\times d_{model}}$.

Multi-head self attention performs extremely well in capturing the long-range dependency. Thus many computer vision tasks, such as image classification\cite{bello2020attention} and object detection\cite{wang2018nonlocal}, employ self-attention as add-ons to boost the performance of traditional CNNs. Here we borrow this idea to further extract text features. $\Mat{Z} \in \mathbb{R}^{l \times d_{model}}$ is treated as the input of CNN, where $l$ is the length of the post (padded or truncated if necessary), and $d_{model}$ is the dimension of the representation vector of each word.

A convolution operation is a filter $\textbf{\emph{w}} \in R^{h\times d_{model}}$ applied to a window of $h$ words to extract high-level features, where $h$ is called the reception field involving $h$ words simultaneously.
For example, feature vector $\textbf{\emph{v}}_{i}$ can be generated from a window of words $\textbf{\emph{z}}_{i:i+h-1}$:
\begin{equation}\label{eq:7}
    \textbf{\emph{v}}_{i} = \sigma(\textbf{\emph{w}} \cdot \textbf{\emph{z}}_{i:i+h-1} + \textbf{\emph{b}}),
\end{equation}
where $\textbf{\emph{z}}_{i:i+h-1}$ is the concatenation of pre-train word representations, \ie $f_{\text{concatenate}}(z_{i}, ..., z_{i-h+1})$, \textbf{\emph{b}} $\in \mathbb{R}$ is the bias term and $\sigma$ is the activation function like ReLU. In a sentence, the filter will be applied to every possible window of words:
\begin{equation}\label{eq:8}
    \textbf{\emph{v}} = [\textbf{\emph{v}}_{1},\textbf{\emph{v}}_{2},\,...\,,\textbf{\emph{v}}_{n-h+1}].
\end{equation}
Then the max pooling is adopted over the calculated feature map to highlight the most present feature in the patch:
\begin{equation}\label{eq:9}
    \hat{\textbf{\emph{v}}} = f_{\text{max-pooling}}(\textbf{\emph{v}}).
\end{equation}

Filters with different reception field $k$ focus on words in k-gram pattern. Inspired by \cite{kim-2014-convolutional}, we use $n$ filters with various window size, and concatenate the output to serve as the ultimate text representation $\Mat{t}$:
\begin{equation}\label{eq:cnn-concat}
    \Mat{t} = f_{\text{concatenate}}(\hat{\textbf{\emph{v}}}_{1},\hat{\textbf{\emph{v}}}_{2},\,...\,,\hat{\textbf{\emph{v}}}_{n}).
\end{equation}

\subsubsection{Contrastive Learning}
Having derived the propagation representation from the GNN module and the semantic representations from the CNN module, we want to fuse these two heterogeneous representations to uncover their underlying relations and find a more comprehensive portrayal of posts. Although many empirical analyses between content features and spreading patterns are conducted, it is still unclear what features of the content are the critical factors to affect the spreading pattern\cite{2016Physics}.

From a probability perspective, we suppose text content could provide a low entropy prior for propagation structure identification. Towards this end, we devise the self-supervised learning tasks from the perspectives of instance- and cluster-wise discrimination.

\vspace{5pt}
\noindent\textbf{Propagation-Semantic Instance Discrimination (PSID).}
For a given post, the propagation tree and semantic content simultaneously characterize it from two different views.
To model such co-occurring relations, we consider a self-supervised task of post self-discrimination --- predicting whether two views are from the same post instance.
Formally, we treat the views of the same post instances as the positive pairs, \ie $\{(\textbf{\emph{g}}_{i}, \textbf{\emph{t}}_{i})| i \in \Set{C}\}$, and that of any two different post instances as the negative pairs, \ie $\{(\textbf{\emph{g}}_{i}, \textbf{\emph{t}}_{j})| i, j \in \Set{C}, i \neq j\}$. Note that $\textbf{\emph{g}}_{i}$ and $\textbf{\emph{t}}_{j}$ are generated graph and text representation by corresponding encoder $E_{1}$ and $E_{2}$, respectively.
The positive pairs offer the auxiliary supervision to enforce the consistency between two views of the same posts; whereas, the negative pairs encourage the divergence among different posts.
The objective function of post self-discrimination is formulated as:
\begin{equation}\label{eq:12}
    \mathcal{L}_{\text{ssl}} = \sum_{i\in\Set{C}}-\log[\frac{\exp(s(\textbf{\emph{g}}_{i}, \textbf{\emph{t}}_{i})/\tau)}{\sum_{j \in \Set{C}}\exp(s(\textbf{\emph{g}}_{i}, \textbf{\emph{t}}_{j})/\tau)}]
\end{equation}
where $s(\cdot)$ stands for the similarity function, which we set as the inner product; $\tau$ is a temperature hyper-parameter, which plays a critical role in mining hard negatives.

\begin{figure*}[!t]
\centering
\includegraphics[width=\textwidth]{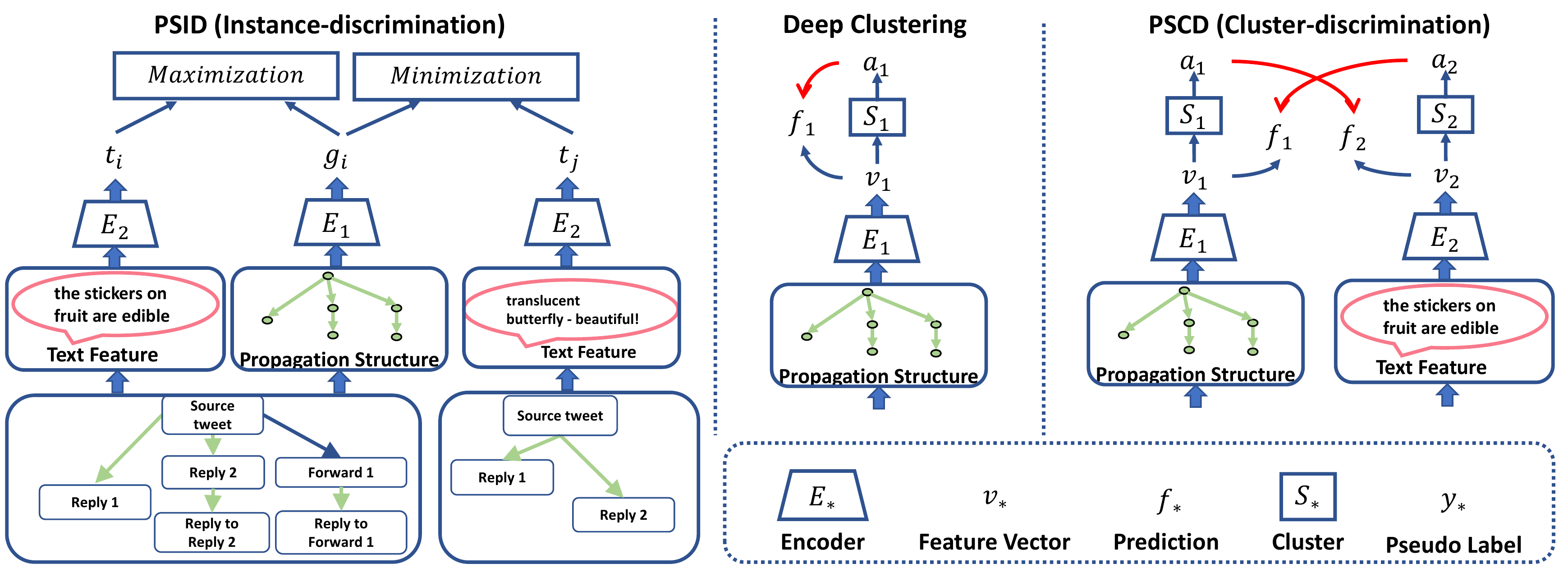}
\caption{Overview of instance discrimination and cluster-discrimination approach. Propagation-Semantic Instance Discrimination (PSID) and Propagation-Semantic Cluster Discrimination (PSCD) are represented. The blue arrows stand for feed forward, while the red arrows stand for back propagation.}
\label{fig_dc}
\end{figure*}

\vspace{5pt}
\noindent\textbf{Propagation-Semantic Cluster Discrimination\\ (PSCD).}
Going beyond the instance-wise discrimination \cite{he2020momentum, Ting2020}, we are motivated by the prior works \cite{caron2019deep} and also devise a cluster-wise discrimination task. Since PSID is a contrastive learning method that need to compute pairwise comparisons, its performance relies on the negative sampling strategy. We aim to utilize simple clustering methods like K-means to cluster the data points and enforce consistency between cluster assignments produced from different views (\ie propagation and semantic) of the same post. Intuitively, similar posts are highly likely to have similar views, thus we can group posts into several clusters and distill cluster-wise features of posts. The self-supervised task can be formulated as maximizing in-group homogeneity. More clearly, the cluster assignment of propagation post would act as pseudo-labels to the semantic representation, and vice versa. Technically, the model consists of two components, encoders $E_{1}$ and $E_{2}$, which map a post $c$'s propagation representation $\Mat{g}$ and semantic representation $\Mat{t}$ into the $d$-dimensional cluster-wise representations, respectively.
Then the discriminator assigns these cluster-wise representations to one of the $K$ clusters:
\begin{gather}
    \min_{\Mat{S}_{1}}\sum_{c\in\Set{C}}\min_{\Mat{a}_{1}}\norm{E_{1}(\Mat{g})-\Mat{S}_{1}\Mat{a}_{1}}_{2}^{2} + \nonumber\\ \min_{\Mat{S}_{2}}\sum_{c\in\Set{C}}\min_{\Mat{a}_{2}}\norm{E_{2}(\Mat{t})-\Mat{S}_{2}\Mat{a}_{2}}_{2}^{2},\nonumber\\
    \text{s.t.}\quad\Trans{\Mat{a}}_{1}\Mat{1}=1,~~\Trans{\Mat{a}}_{2}\Mat{1}=1,
\end{gather}

where $\Mat{S}_{G}$ and $\Mat{S}_{T}\in\Space{R}^{d\times K}$ separately denote the trainable centroid matrices of graph structure and semantic content, whose $k$-th column is the centroid representation of the $k$-th cluster;
$\Mat{a}_{1}$ and $\Mat{a}_{2}\in\{0,1\}^{K}$ are one-hot encoding, which represent the cluster assignments based on graph structure and semantic content, respectively.
Here we adopt the iterative optimization framework: given the centroid matrices, we learn the best cluster indicators as $\Mat{a}_{1}$ and $\Mat{a}_{2}$; then the assignment results guide the optimization of $\Mat{S}_{1}$ and $\Mat{S}_{2}$.
As a result, a set of optimal cluster assignments $\{\Mat{a}^{*}_{1},\Mat{a}^{*}_{2}|c\in\Set{C}\}$ are established as the pseudo-labels or supervision signal to enhance the representations of posts, as follows:
\begin{gather}
    \Lapl_{\text{ssl}} = \sum_{c\in\Set{C}}l(f_{1}(E_{1}(\Mat{g})),\Mat{a}_{2}) + l(f_{2}(E_{2}(\Mat{t})),\Mat{a}_{1}),
\end{gather}
where $l(\cdot)$ is the negative log-softmax function; $f_{1}(\cdot)$ is a trainable classifier, which takes the swapped propagation representation $\Mat{g}$ as input to predict the semantic-aware cluster assignment $\Mat{a}_{2}$, analogously to $f_{2}(\cdot)$.
The procedure of PSCD and PSID are shown in Figure \ref{fig_dc} and Algorithm \ref{alg1}.

\begin{algorithm}[!t]
\caption{Training Procedure} 
\label{alg1}
\SetKwInOut{Input}{Input}
\SetKwInOut{Output}{Output}
\SetKwComment{Comment}{// }{}
\KwIn{Attributed graph with node features, edges and labels $\mathcal{G}=\{\mathcal{V}, \mathcal{E}, \mathcal{X},\mathcal{Y}\}$, Number of GNN layers, number of CNN layers, batches and training epochs $G_{L}, C_{L}, B, E$} 
\KwOut{Classification Result: Rumor or Non-Rumor}
Initialization: $h_v^0 \leftarrow x_v$\;
\For{e=1, ..., E}
{
    \For{b=1, ..., B}
    {
        \For{l=1, ..., $G_{L}$}
        {
            $g_{v} \leftarrow Eq.(2)(3), \forall v \in \mathcal{V}_b$\Comment*[r]{Propagation representation}
        }
        \For{l=1, ..., $C_{L}$}
        {
            $z_{x} \leftarrow Eq.(4)(5)(6), \forall x \in \mathcal{X}$\Comment*[r]{Semantic Representation}
        }
    }
    \If{PSCD}{
    freeze S and get a $\leftarrow Eq.(11)$\;
    freeze a and update S $\leftarrow Eq.(11)$\Comment*[r]{K-means}
    $\mathcal{L}_{ssl} \leftarrow Eq.(12)$\;
    }
    \If{PSID}{
    negative sampling\;
    $\mathcal{L}_{ssl} \leftarrow Eq.(10)$\;
    }
    $\mathcal{L}_{main} \leftarrow Eq.(14)$\;
    $\mathcal{L} \leftarrow Eq.(15)$\;
}
\end{algorithm}

\subsubsection{Rumor Prediction}
We build a rumor detector model on the top of a post $c$'s propagation representation $\Mat{g}$:
\begin{gather}
p(c) = \sigma(\Mat{W}\Mat{g}+\Mat{b}),
\end{gather}
where $p(c)$ presents the probability of the post being the rumor;
$\Mat{W}$ is the trainable matrix, and $\Mat{b}$ is the bias term.
This term is used in a cross-entropy loss, framing the main supervised learning task:
\begin{gather}
    \Lapl_{\text{main}}=-\sum_{c\in\Set{C}}y\log(p(c)),
\end{gather}
where $y$ is the binary label indicating whether a post is rumor.
We adopt a multi-task learning strategy to jointly optimize the main supervised learning task and the auxiliary self-supervised learning task:
\begin{gather}
    \Lapl=\Lapl_{\text{main}} + \lambda\Lapl_{\text{ssl}},
\end{gather}
where $\lambda$ is a hyper-parameter to control the strength of the auxiliary task.
\section{Experiments and Analyses}
We perform experiments on three real-world datasets to evaluate our proposed method, especially the self-supervised learning module. We aim to answer the following research questions:
\begin{itemize}
    \item RQ1: How does SRD perform as compared with state-of-the-art methods?
    \item RQ2: How effective are Graph Neural Network, Text Encoder and SSL, respectively, in improving the rumor detection performance of SRD? And how do different hyper-parameter settings (e.g., loss scalar $\lambda$, temperature parameter $\tau$) affect our method?
    \item RQ3: Can SRD perform well on early rumor detection?
\end{itemize}

\subsection{Dataset}
We conduct extensive experiments on three widely used benchmark on rumor detection: Twitter\cite{Ma2017}, Weibo\cite{Ma2016}, and PHEME\cite{Zubiaga2016}. These datasets are representative since they contain blogs and posts from the most influential social media sites in China and United States, respectively. The Weibo dataset contains two binary labels: \textit{F} stands for False Rumor and \textit{T} stands for True Rumor, while Twitter dataset and PHEME dataset contain multiple labels: \textit{N} stands for Non-rumor, \textit{F} stands for False Rumor, \textit{T} stands for True Rumor, and \textit{U} stands for Unverified Rumor. The rumors from Weibo dataset are from the Sina community management center, which reports various misinformation; while non-rumor events are the posts that are not reported as rumors \cite{Ma2016}. And the labels in Twitter are converted from binary to quaternary according to the veracity tag of the article in rumor debunking websites (e.g., snopes.com, Emergent.info, etc\cite{Ma2017}. The PHEME dataset contains tweets about 9 topics that are closely related to politics and people’s livelihood, and the labels are annotated by news practitioners. Following \cite{Bian2020,Ma2018}, we use the same preprocessed dataset: every sample is a propagation tree, whose nodes and edges are posts and their retweet or response relationships. The statistics of the datasets are summarized in Table \ref{table:1}, nonexistent or unknown statistics are marked as "-".

\begin{table}[!t]
\renewcommand{\arraystretch}{1.3}
\caption{Statistics of the Dataset}
\label{table:1}
\centering
\resizebox{\columnwidth}{!}{
\begin{tabular}{l|c c c}
 \hline
 Statistic & \textit{Weibo} & \textit{Twitter} & \textit{PHEME}\\ 
 \hline\hline
 \# of posts & 3,805,656 & 204,820 & 33288\\ 
 \# of users & 2,746,818 & 173,487 & -\\ 
 \# of events & 4,664 & 818 & 6425\\\hline
 \# of True Rumors & 2,351 & 205 & 1067\\
 \# of False Rumors & 2,313 & 205 & 638\\
 \# of Unverified Rumors & - & 203 & 698\\
 \# of Non-rumors & - & 205 & 4022\\\hline
\end{tabular}}
\end{table}

\subsection{Baseline}
We compared our proposed method with the following strong baselines, including classic and state-of-the-art rumor detection methods.
\begin{itemize}
\item \textbf{DTC}~\cite{Castillo2011}: A method extracting features manually, based on which they train a Decision Tree Classifier to predict the label of events.
\item \textbf{SVM-RBF}~\cite{Yang2012}: SVM Classifier with RBF kernel, using the hand-crafted feature to do classification.
\item \textbf{SVM-TS}~\cite{Ma2015}: SVM based model that utilizes time series information to capture temporal features for prediction.
\item \textbf{SVM-TK}~\cite{Ma2017}: SVM Classifier takes propagation structure into consideration by designing and leveraging Propagation Tree Kernel.
\item \textbf{RvNN}~\cite{Ma2018}: A method that employs Recursive Neural Networks with GRU units, integrates both propagation structure and content semantics for detecting rumors from microblog posts.
\item \textbf{STS-NN}~\cite{huang2020deep}: A method that treats the spatial structure and the temporal structure as a whole to model the message propagation.
\item \textbf{PPC}~\cite{Liu2018}: A method that focuses on user characteristics to do propagation path classification with a combination of recurrent and convolutional networks.
\item \textbf{BiGCN}~\cite{Bian2020}: The first study to solve rumor detection problem with GCN, focusing on both rumor propagation and rumor dispersion with top-down propagation tree and bottom-up propagation tree, respectively.
\item \textbf{EBGCN}~\cite{wei2021towards}: The first study to explore propagation uncertainty for rumor detection, which adaptively rethinks the reliability of latent relations
by adopting a Bayesian approach.
\end{itemize}

\subsubsection{Implementation Details and Metrics}
We focus on comparison with the state-of-the-art model, so the other baseline models' experiment results are cited from \cite{Bian2020}. We implement BiGCN, EBGCN and our proposed method with Pytorch, and all the three models are tested on Geforce RTX 2080Ti. When running EBGCN on Weibo dataset, it raises an "Out of Memory" issue. To make a fair comparison, we randomly split the dataset into 5 parts with a fixed seed, and employ 5-fold validation to get the robust result. We utilize Adam algorithm \cite{Diederik2014} as optimizer with default setting unchanged (0.9 and 0.999, respectively). Mini-batch size is set to 128. The learning rate varies during training with cosine annealing strategy \cite{loshchilov2017sgdr}. For semantic modules, rare word deletion is adopted to reduce noise in data, \ie deleting words that appear less than 2 times. Note that this operation could lead to the emptiness of samples, which should also be removed. Text words are truncated or padded to a fixed length L to form the input matrix. The convolutional kernel size in CNN is set to (3,4,5) with 100 feature maps each, and a dropout rate of 0.5. Word embeddings that act as the input of the text encoder are initialized with 300-dimensional word2vec\cite{10.5555/2999792.2999959} trained on social media corpora, where words not seen before would be initialized from a uniform distribution. We also try to inference Bert\cite{devlin-etal-2019-bert} to acquire the representation of sentences, which indicates that the choice of pretrained embedding is not crucial in our method. We keep the word vectors trainable during the process, for propagation information may leverage constraints on them. All of the trainable parameters are initialized with Xavier initializer\cite{Xavier2010}. For other hyper-parameters, we applied grid search: the start learning rate is chosen between \{0.0005, 0.001, 0.01, 0.05\}, and the temperature is searched through \{0.1, 0.2, 0.5, 1, 2\}; $\lambda$, which balance the loss of prime learning $\mathcal{L}_{Prime}$ and that of contrastive learning $\mathcal{L}_{cl}$, is tuned within the range of \{0.005, 0.01, 0.05, 0.1\}.

For fair comparison, We follow the previous work \cite{Bian2020,Liu2018} to adopt Accuracy, Precision, Recall and F1 score as indicators to evaluate the overall performance of models.

\begin{table}[!t]
\renewcommand{\arraystretch}{1.3}
\caption{Performance Comparison on Twitter Dataset}
\label{table:2}
\centering
\resizebox{0.45\textwidth}{!}{
\begin{tabular}{c|c|c c c c}
 \hline
 \multirow{2}*{Method} & \multirow{2}*{Accuracy} & N & F & T & U \\ 
 \cline{3-6}
 ~ & ~ & F1 & F1 & F1 & F1 \\
 \hline\hline
 DTC & 0.473 & 0.254 & 0.080 & 0.190 & 0.482 \\

 SVM-RBF & 0.553 & 0.670 & 0.085 & 0.117 & 0.361 \\

 SVM-TS & 0.574 & 0.755 & 0.420 & 0.571 & 0.526 \\

 SVM-TK & 0.732 & 0.740 & 0.709 & 0.836 & 0.686 \\

 RvNN & 0.737 & 0.662 & 0.743 & 0.835 & 0.708 \\

 STS-NN & 0.810 & 0.753 & 0.766 & 0.890 & 0.838 \\
 
 PPC & 0.863 & 0.820 & 0.898 & 0.843 & 0.837 \\
 
 EBGCN & 0.871 & 0.820 & 0.865 & 0.922 & 0.861 \\

 Bi-GCN & 0.886 & 0.830 & 0.881 & 0.942 & 0.885 \\
 \hline\hline
 \textbf{SRD-PSCD} & 0.891 & 0.837 & 0.889 & 0.945 & 0.898 \\
 \textbf{SRD-PSID} & \textbf{0.905} & \textbf{0.857} & \textbf{0.906} & \textbf{0.953} & \textbf{0.909} \\
 \hline
\end{tabular}}
\end{table}

\begin{center}
\begin{table}[!t]
\renewcommand{\arraystretch}{1.3}
\caption{Performance Comparison on Weibo Dataset}
\label{table:3}
\centering
\resizebox{0.45\textwidth}{!}{
\begin{tabular}{c|c|c|c|c|c}
 \hline
 Method & Class & Accuracy & Precision & Recall & F1 score \\ 
 \hline
 \multirow{2}*{DTC} & F & \multirow{2}*{0.831} & 0.847 & 0.815 & 0.831 \\
 ~ & T & ~ & 0.815 & 0.824 & 0.819 \\
 \hline
 \multirow{2}*{SVM-RBF} & F & \multirow{2}*{0.879} & 0.777 & 0.656 & 0.708 \\
 ~ & T & ~ & 0.579 & 0.708 & 0.615 \\
 \hline
 \multirow{2}*{SVM-TS} & F & \multirow{2}*{0.885} & 0.950 & 0.932 & 0.938 \\
 ~ & T & ~ & 0.124 & 0.047 & 0.059 \\
 \hline
 \multirow{2}*{RvNN} & F & \multirow{2}*{0.908} & 0.912 & 0.897 & 0.905 \\
 ~ & T & ~ & 0.904 & 0.918 & 0.911 \\
 \hline
 \multirow{2}*{STS-NN} & F & \multirow{2}*{0.912} & 0.912 & 0.912 & 0.908 \\
  ~ & T & ~ & 0.911 & 0.915 & 0.913 \\
 \hline
 \multirow{2}*{PPC} & F & \multirow{2}*{0.916} & 0.884 & 0.957 & 0.919 \\
 ~ & T & ~ & 0.955 & 0.876 & 0.913 \\
 \hline
 \multirow{2}*{Bi-GCN} & F & \multirow{2}*{0.947} & \textbf{0.972} & 0.921 & 0.946 \\
 ~ & T & ~ & 0.925 & \textbf{0.974} & 0.949 \\
 \hline
  \multirow{2}*{\textbf{SRD-PSCD}} & F & \multirow{2}*{0.949} & 0.954 & 0.945 & 0.949 \\
 ~ & T & ~ & 0.945 & 0.954 & 0.950 \\
 \hline
   \multirow{2}*{\textbf{SRD-PSID}} & F & \multirow{2}*{\textbf{0.962}} & 0.967 & \textbf{0.956} & \textbf{0.961} \\
 ~ & T & ~ & \textbf{0.956} & 0.967 & \textbf{0.962} \\
 \hline
\end{tabular}
}
\end{table}
\end{center}

\begin{center}
\begin{table}[!t]
\renewcommand{\arraystretch}{1.3}
\caption{Performance Comparison on PHEME Dataset}
\label{table:4}
\centering
\resizebox{0.45\textwidth}{!}{
\begin{tabular}{c|c|c c c c}
 \hline
 \multirow{2}*{Method} & \multirow{2}*{Accuracy} & N & F & T & U \\ 
 \cline{3-6}
 ~ & ~ & F1 & F1 & F1 & F1 \\
 \hline\hline
 EBGCN & 0.819 & 0.895 & 0.712 & 0.702 & 0.561 \\
 Bi-GCN & 0.833 & 0.904 & 0.731 & 0.723 & 0.604 \\

 \textbf{SRD-PSCD} & \textbf{0.839} & \textbf{0.905} & 0.721 & \textbf{0.741} & \textbf{0.605} \\

 \textbf{SRD-PSID} & 0.838 & \textbf{0.905} & \textbf{0.774} & 0.734 & 0.604 \\
 \hline
\end{tabular}}
\end{table}
\end{center}
\vspace{-80pt}

\subsection{Result Analysis}
To answer \textbf{RQ1}, we compare SRD with the algorithms introduced in Section 4.2, results on Twitter, Weibo and PHEME dataset are shown in Table \ref{table:2}, \ref{table:3} and \ref{table:4} respectively. We can observe:

SRD consistently outperforms all of the baselines on all datasets. And as we can see, deep learning models achieve a much better result than models based on hand-crafted features, which is normal because they enable automatic feature engineering and high-level representation. What's more, our proposed method and Bi-GCN perform better than RvNN and PPC (RNN + CNN), which proves the importance of research in Graph Neural Networks for rumor detection.
The state-of-the-art method of rumor detection, the performances of Bi-GCN, EBGCN and STS-NN are less satisfactory compared to our proposed method PSCD and PSID in all of the measurement metrics. Hereby, we attribute our more feasible solution to two following reasons: 1) BiGCN and EBGCN do not incorporate linguistic features or graph structures in post information, instead, they focus more on feature mining in rumor propagation and rumor dispersion; however, the performance of GCN tends to decrease as the number of nodes in propagation tree get smaller because the information provided become sparser. 2) Our proposed method incorporates SSL in the framework, this auxiliary task could improve the performance of the primary task learning, and the text content can provide supplementary information and constraints to the Graph Neural Network branch. Ablation study in Section 4.4 further proves the opinion.
The PSID version of SRD performs better than or comparable to PSCD version on three datasets, which demonstrates the effectiveness of introducing negative sampling and pair comparison to boost the performance. The multi-view strategy enables the instance-discrimination approach to benefit from the additional information from the other views, while the cluster-discrimination approach focuses more on the homogeneity of views of the same event.

\begin{center}
\begin{table}[!t]
\renewcommand{\arraystretch}{1.3}
\caption{Ablation study of SRD demonstrated the advantage of the GCN, text encoder and contrastive learning on Twitter Dataset.}
\label{table:5}
\centering
\resizebox{0.45\textwidth}{!}{
\begin{tabular}{c|c|c c c c}
 \hline
 \multirow{2}*{Method} & \multirow{2}*{Accuracy} & N & F & T & U \\ 
 \cline{3-6}
 ~ & ~ & F1 & F1 & F1 & F1 \\
 \hline
 SRD-TEXT & 0.803 & 0.772 & 0.768 & 0.885 & 0.783 \\
 \hline
 SRD-GRAPH & 0.880 & 0.825 & 0.901 & 0.924 & 0.877 \\
 \hline
 Bi-GCN & 0.886 & 0.830 & 0.881 & 0.942 & 0.885 \\
 \hline
 SRD-CONCAT & 0.888 & 0.826 & \textbf{0.910} & 0.940 & 0.860 \\
 \hline
 \textbf{SRD-PSID} & \textbf{0.905} & \textbf{0.857} & 0.906 & \textbf{0.953} & \textbf{0.909} \\
 \hline
\end{tabular}}
\end{table}
\end{center}

\vspace{-60pt}
\subsection{Study of SRD (RQ2)}
\subsubsection{Ablation Analysis}
To get deeper insights into the different components (especially the self-supervised learning module) of our method, we fully investigate their impacts. \\
\textbf{SRD-TEXT} is a variant of SRD, which only uses rumor texts to do prediction. \\
\textbf{SRD-GRAPH} is a variant of SRD, which only utilizes graph propagation structure and a 2-layer GCN to do prediction. \\
\textbf{SRD-CONCAT} is a variant of SRD, which incorporates both of the structured information and semantic features in the framework, but directly concatenates them without adding self-supervised learning module.

Table \ref{table:5} summarizes the comparison results. It can be observed that:
The Text-only method is poor performing compared to all the other methods, which shows the importance of modeling the relations between entities, as solely relying on linguistic features to detect rumors can be unstable.
BiGCN's performance is better than SRD-GRAPH, the possible reason is that they incorporate the rumor dispersion by adding bottom-up propagation tree. 
SRD-CONCAT performs better than those methods using either one of text contents or graph feature. This indicates that linguistic features can provide complementary information to propagation patterns, thereby improving the detection results. 
SRD-PSID outperforms SRD-CONCAT, which supports our opinion that self-supervised learning can better utilize heterogeneous features than simple concatenation.
According to Table \ref{table:propagation}, the column "\% of P and !L" is the percentage of the instances which are correctly classified with propagation information rather than linguistic information, analogously to another column. From this table, we can further conclude that the effectiveness of both taking propagation pattern and linguistic features into consideration.


\begin{center}
\begin{table}[!t]
\renewcommand{\arraystretch}{1.3}
\caption{Effects of the Network Depth}
\label{table:6}
\centering
\resizebox{\columnwidth}{!}{
\begin{tabular}{c|c|c|c|c|c}
 \hline
 Dataset & Metric & 1 & 2 & 3 & 4 \\ 
 \hline
 \multirow{5}*{Twitter} & Accuracy & 0.860 & 0.905 & 0.879 & 0.858 \\
 ~ & NF1 & 0.760 & 0.857 & 0.818 & 0.774 \\
 ~ & FF1 & 0.837 & 0.906 & 0.882 & 0.829 \\
 ~ & TF1 & 0.905 & 0.953 & 0.921 & 0.896 \\
 ~ & UF1 & 0.794 & 0.909 & 0.896 & 0.878 \\
 \hline
 \multirow{3}*{Weibo} & Accuracy & 0.958 & 0.962 & 0.955 & 0.956 \\
 ~ & NF1 & 0.958 & 0.961 & 0.944 & 0.956 \\
 ~ & FF1 & 0.958 & 0.962 & 0.958 & 0.956 \\ \hline
\end{tabular}
}
\end{table}
\end{center}

\begin{center}
\begin{table}[!t]
\renewcommand{\arraystretch}{1.3}
\caption{Effectiveness of propagation structure}
\label{table:propagation}
\centering
\resizebox{0.45\textwidth}{!}{
\begin{tabular}{c|c c c c c}
 \hline
 Dataset & P Acc & L Acc & \% of P and !L & \% of L and !P\\ 
 \hline
 Twitter & 0.898 & 0.803 & 11.25 & 2.50 \\
 Weibo & 0.954 & 0.910 & 8.125 & 3.75 \\
 \hline
\end{tabular}}
\end{table}
\end{center}

\vspace{-80pt}
\subsubsection{Effects of Network Depth}

We vary the network depth $L$ to investigate the efficiency of the usage of multiple embedding propagation layers. In more detail, the layer number is tested within the range of \{1,\,2,\,3,\,4\}. We use SRD-1 to represent 1 layer variant, and similar notations for other variants. Figure \ref{table:6} summarize the result of comparison. We have the following observation:

Clearly, SRD achieves the best result at 2 layers. In the case of network length $L \geq 3$, the model performance begins to decrease, which is consistent with the choice of \cite{Bian2020}. We think the possible reason might be that high-order relations between entities, like third- and fourth-order connectivities, are rare in the dataset; as seen in Figure \ref{fig_fa}, almost 65 percent of nodes are first-order connections. Thus 3 or more hop neighborhoods contribute little information to the training. Another reason is that Graph Convolution Network suffers more from over-smoothing as layer depth increasing\cite{li2018deeper}.
Analyzing Table \ref{table:2}, Table \ref{table:3} and Table \ref{table:6} together, we can find that SRD outperforms baselines which are not based on GCN. It again proves the advantages of taking the rumor propagation pattern into consideration.

\subsubsection{Effects of Loss Scalar}

\begin{figure}[!t]
\centering
\includegraphics[width=\columnwidth]{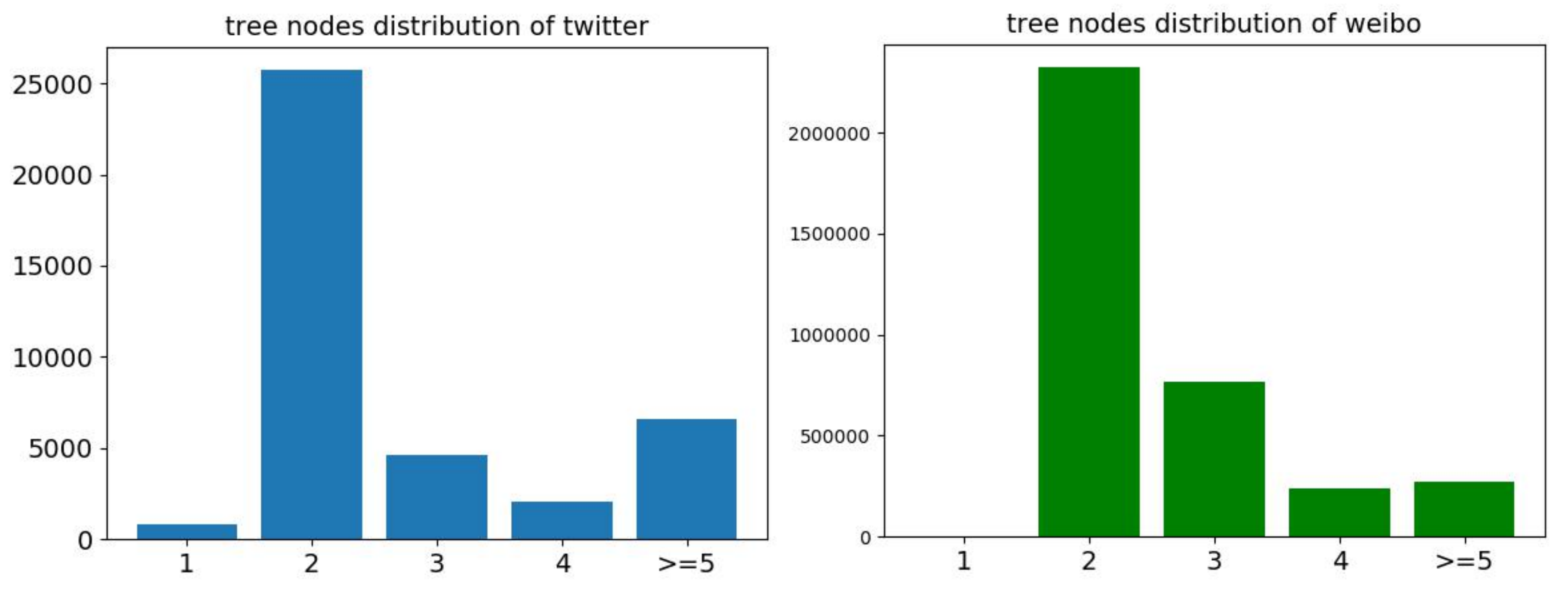}
\caption{Depth of Propagation Tree Nodes Distribution on Twitter and Weibo. $x$ axis means the depth of nodes in a propagation tree; e.g. 1 denotes the root node, then 2 will represent the direct children of root node. For better visualization, we classify those whose depth are greater than 5 into a single category.}
\label{fig_fa}
\end{figure}

As we adopt the \textit{primary and auxiliary} framework, $\lambda$ should be tuned carefully to avoid the negative influence from the supplementary task when propagating gradient. We search for the near optimal value within a small interval starting from 0. Fig \ref{fig_fa} illustrates the performance of SRD with five $\lambda$ values \{0, 0.005, 0.01, 0.05, 0.1\}. As seen from the figure, the performance of SRD changes smoothly near the peak in Weibo dataset and PHEME dataset. It varies more in Twitter dataset, we suppose the possible reason is that the scale of the dataset is relatively small (seen in Table \ref{table:1}). In most cases, SRD is not sensitive to hyperparameter $\lambda$. Another observation is that the performance rises with the increase of the values of $\lambda$ at the beginning, after reaching the optimal value at around 0.01, it keeps unchanged or decreases in a small value. We also run test with $\lambda$ equaling to 0.5, only to observe a significant drop in model performance. According to the above observation, we can draw a conclusion that auxiliary task with a small $\lambda$ can improve the primary task while that with a large one might harm or even mislead the supervised task.

\begin{figure}[!t]
\centering
\includegraphics[width=0.7 \columnwidth]{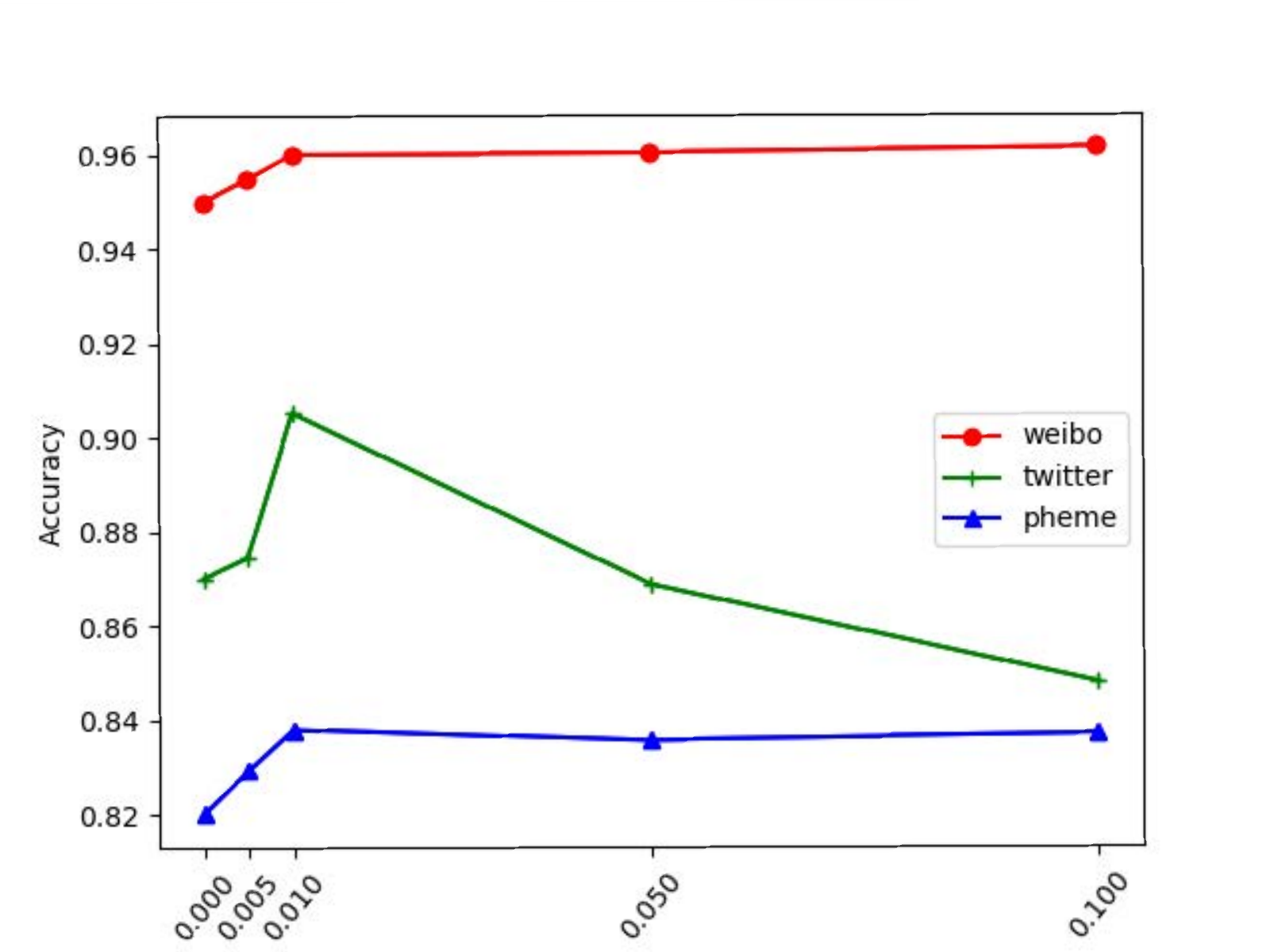}
\caption{Model Accuracy \textbf{w.r.t} loss scalar $\lambda$}
\label{fig_lam}
\end{figure}

\subsubsection{Effects of Temperature Parameter}

\begin{figure}[!t]
\centering
\includegraphics[width=\columnwidth]{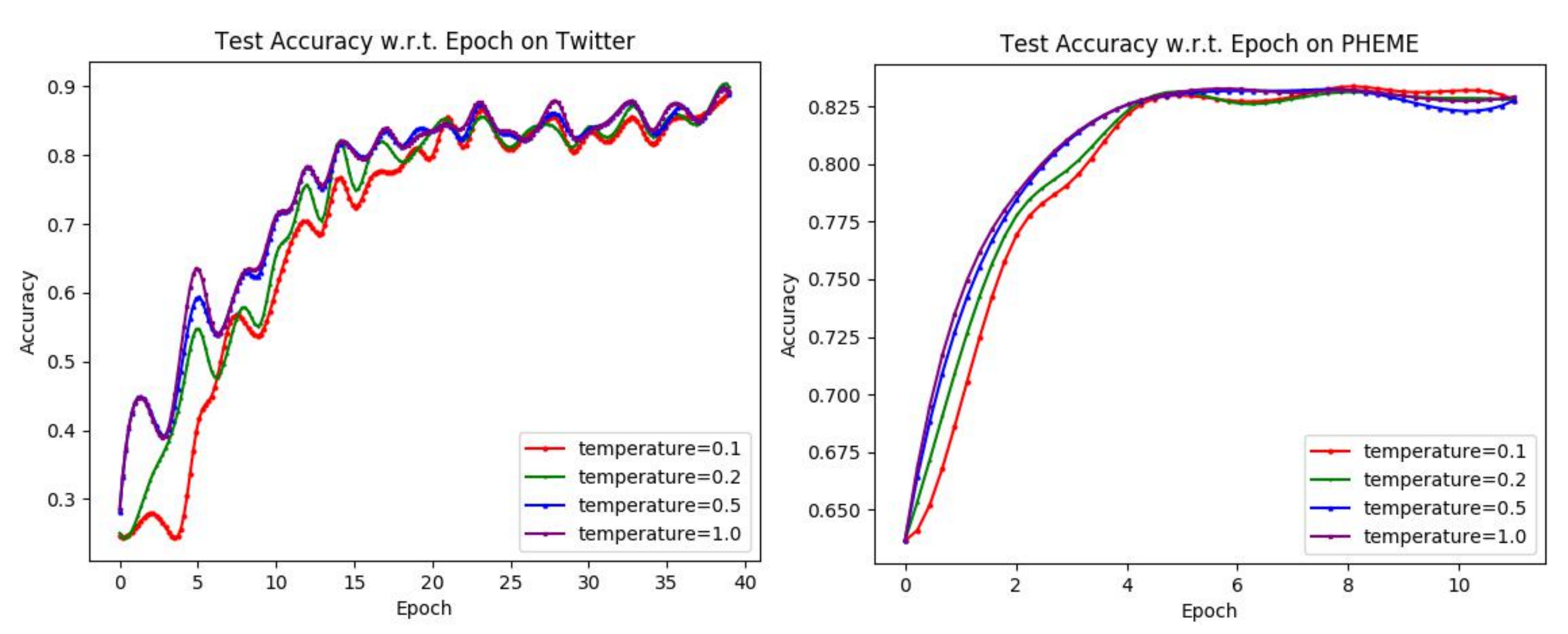}
\caption{Model Accuracy \textbf{w.r.t} temperature parameter $\tau$}
\label{fig_temp}
\end{figure}

According to \cite{Ting2020, wang2021understanding}, temperature parameter $\tau$ plays an important role in hard negative sampling and an appropriate $\tau$ could benefit the representation learning task when cosine similarity is hired as similarity function. We leverage both \textit{cosine similarity} and \textit{dot product} as our similarity measurement, finding that dot product works better in our task. Fig \ref{fig_temp} shows the test accuracy with respect to training epochs on Twitter dataset and PHEME dataset. Weibo dataset converges in three epochs, so it is excluded from the figure. We can observe that: 1) SRD is not sensitive to the value of $\tau$ since we adopt dot product rather than cosine similarity, it is consistent with the result in \cite{Ting2020}: without $l_{2}$ normalization, model with different $\tau$ roughly converges to the same point. 2) As the value of $\tau$ increasing, the model converges faster, we attribute this phenomenon to the \textit{Uniformity-Tolerance Dilemma}\cite{wang2021understanding}: if the text representation and the propagation pattern from different samples are similar, they are possibly a positive pair but labeled as hard negative mistakenly. This inspires us with the future work of adopting the cosine similarity function with an increasing $\tau$.
\subsection{Early Rumor Detection (RQ3)}
\begin{figure}[!t]
\centering
\includegraphics[width=\columnwidth]{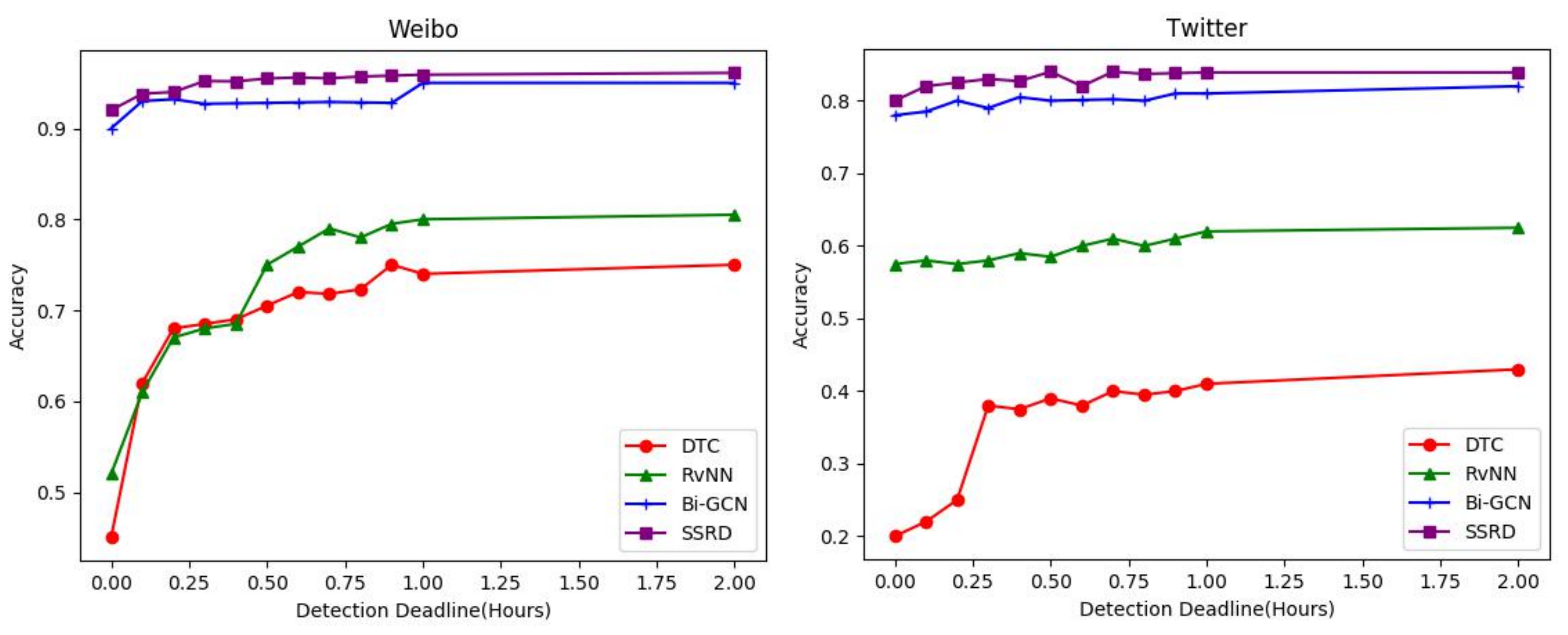}
\caption{Early Detection Performance on Weibo Dataset and Twitter Dataset}
\label{fig_early}
\end{figure}

It is crucial but challenging to detect rumor before it flies high. Many researchers have addressed this issue. Farinneya et al.\cite{farinneya2021active} design an Active Transfer Learning (ATL) strategy to identify rumors with a limited amount of annotated data. Xia et al.\cite{xia2020state} propose a state-independent and time-evolving
Network which captures the event’s unique features in different states. At the early stage, there is limited propagation information to utilize. Thus we expect the text content could help with the issue. To answer \textbf{RQ3}, we conduct experiments on early rumor detection. Fig \ref{fig_early} summarizes the comparison result between our proposed method and 3 chosen baselines DTC, RvNN and Bi-GCN. We have the following observation:

For DTC and RvNN, the performance decreases significantly when the detection deadline is restricted to 2 hours, while SRD and Bi-GCN are not affected. 
Our proposed method is superior to all of the baselines at each deadline, we suppose this gap can be partly attributed to text content representation, which is available once the event begins.

Our proposed method can achieve high performance within half an hour since the release of the source post, which demonstrates that our proposed method can effectively distinguish rumors at a very early stage.

\subsection{Case Study}

\begin{figure}[!t]
\centering
\includegraphics[width=\columnwidth]{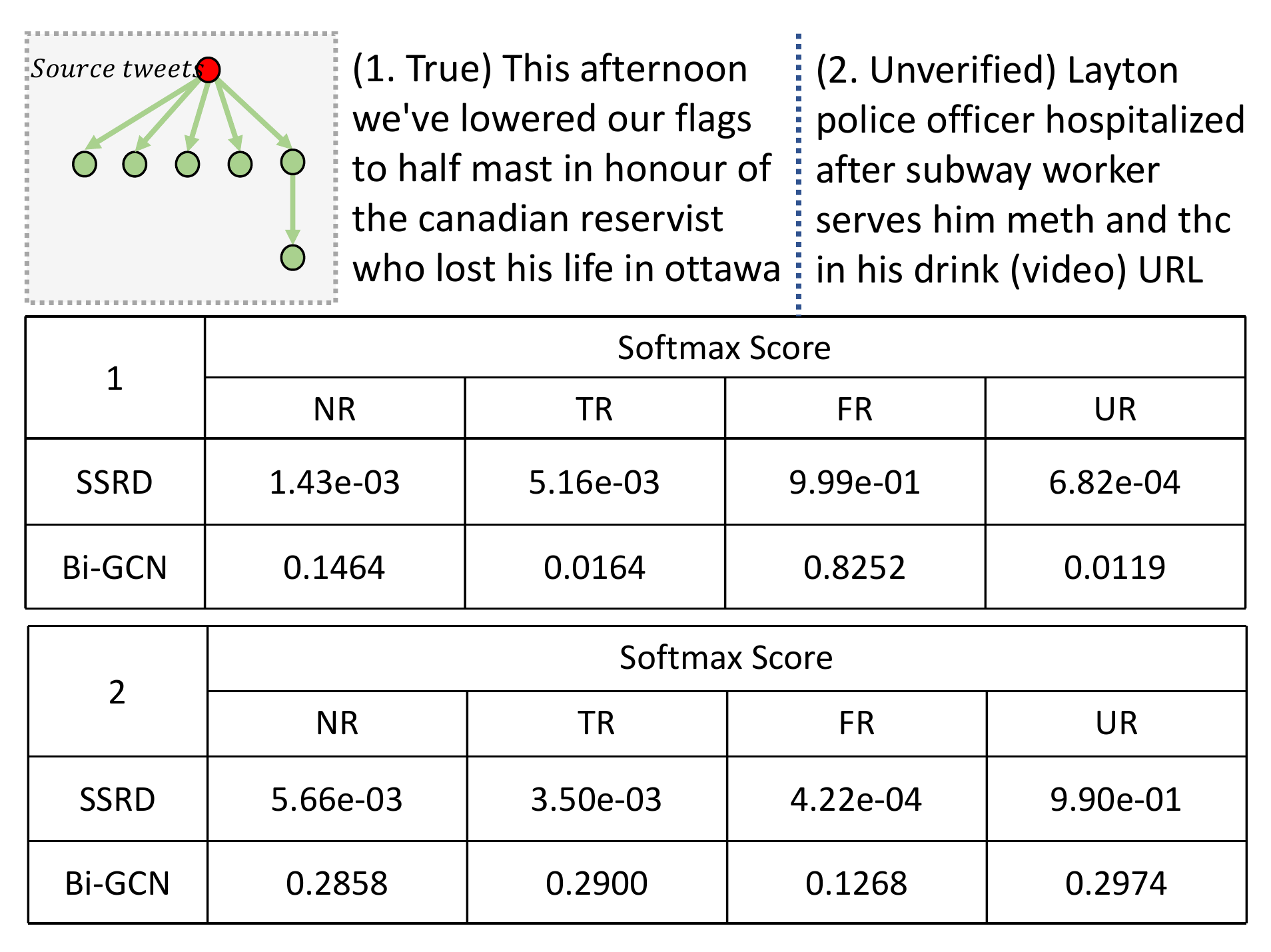}
\caption{Result comparison on two real examples in Twitter. These two examples have the same propagation structure with different ground truth, true rumors and unverified rumors. Softmax scores of prediction results by SRD and Bi-GCN are listed in charts, respectively.}
\label{fig_cs}
\end{figure}

To further validate the importance of textual information, we randomly choose a propagation pattern (e.g. $1\rightarrow5\rightarrow1$), and see if SRD and BiGCN are able to distinguish between 6 examples sharing this same pattern. Within 6 samples, 2 are true rumors (T) and 4 are unverified rumors (U), note that it's a weak indication that propagation structure to some extent is related to categories of posts. Test runs are performed with SRD and BiGCN, and we summarize the comparison result in Fig \ref{fig_cs}. There are two key observations:

Both models can distinguish true rumors and unverified rumors, for they can tell that the post is not T/U when the post belongs to U/T by assigning the lowest probability. 

SRD predicts the label of a given sample at a quite high confidence level of over 0.99, while the highest softmax score of Bi-GCN, which represents the probability, are 0.8252 and 0.2974 respectively. Especially in the unverified case, Bi-GCN can only tell that this post is not a true rumor, and the scores of the other 3 classes are similar. From this perspective, SRD is superior to Bi-GCN, which proves the effectiveness of incorporating supplementary information into the primary graph classification task.

\section{Conclusion and Future Work}
In this work, we proposed a novel method, termed self-supervised learning on rumor detection (SRD), which accounts for both propagation structure and semantic patterns of rumor. By introducing self-supervised learning, we explicitly inject supplementary text representation into the primary rumor detection, which is a supervised graph classification task. In addition, we adopt instance-wise discrimination and cluster-wise discrimination, i.e. NT-Xent loss and XDC to handle the data heterogeneity. Extensive experiments on three real-world datasets demonstrate that SRD performs better than state-of-the-art baselines, hence proving the effectiveness and rationality of incorporating both propagation tree and semantics in the framework. Since our model utilizes text content, which is available in the early stage of an event, it can detect rumors automatically earlier before state-of-the-art baselines. It is of great necessity since debunking rumors early could avoid the influence from becoming irreversible.

Since the propagation tree is permutation invariant, i.e. children of the same node are treated equally without considering timestamp, in the future, we would further improve SRD by incorporating temporal structures. Another potential direction is to join propagation trees and prepare a big graph for semi-supervised prediction, so that we can generalize SRD from inductive to transductive settings and utilize richer adjacency information. We are also interested in altering GCN encoders. Hyperbolic graph neural network\cite{chami2019hyperbolic} could be an ideal one, for its 
advantages in dealing with hierarchical social networks. Another substitution is GCN along with adversarial training like CGAT \cite{ijcai2020-197}, which may be beneficial to model robustness. Also with the development of XAI, we believe graph neural network explainability \cite{DIR, wang2021towards, igv} holds great promise in propagation-based rumor detection.

\bibliographystyle{fcs}
\bibliography{fcs}

\begin{biography}{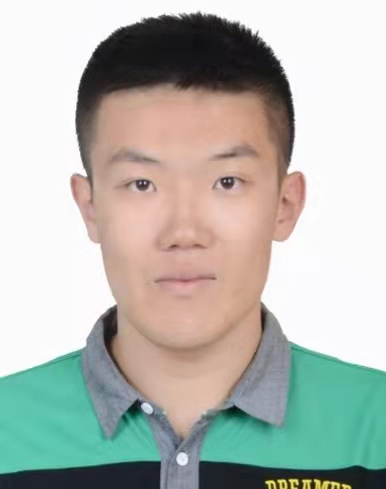}
Yuan Gao received the M.S. degree in Electrical and Computer Engineering, University of Michigan, Ann Arbor, in 2019. He is now a Ph.D. student in the School of Cyberspace Science and Technology at the University of Science and Technology of China (USTC). His research interest lies in fraud detection, representation learning, and graph learning.
\end{biography}

\begin{biography}{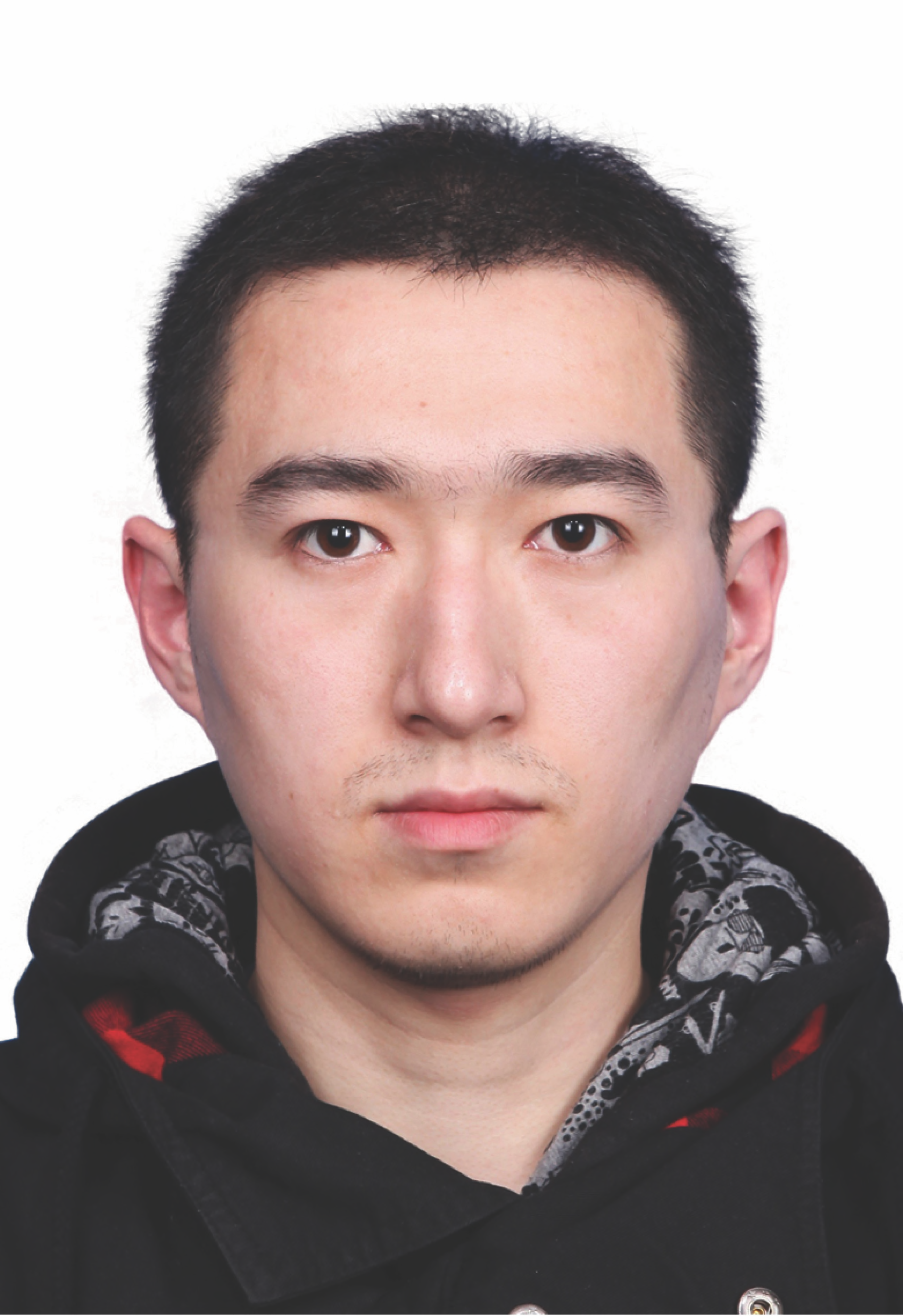}
Xiang Wang is now a professor at the University of Science and Technology of China (USTC). He received his Ph.D. degree from National University of Singapore in 2019. His research interests include recommender systems, graph learning, AI explainability, and AI security. He has published some academic papers on international conferences such as NeurIPS, ICLR, KDD, WWW, SIGIR, and AAAI. He serves as a program committee member for several top conferences such as SIGIR and WWW.
\end{biography}

\begin{biography}{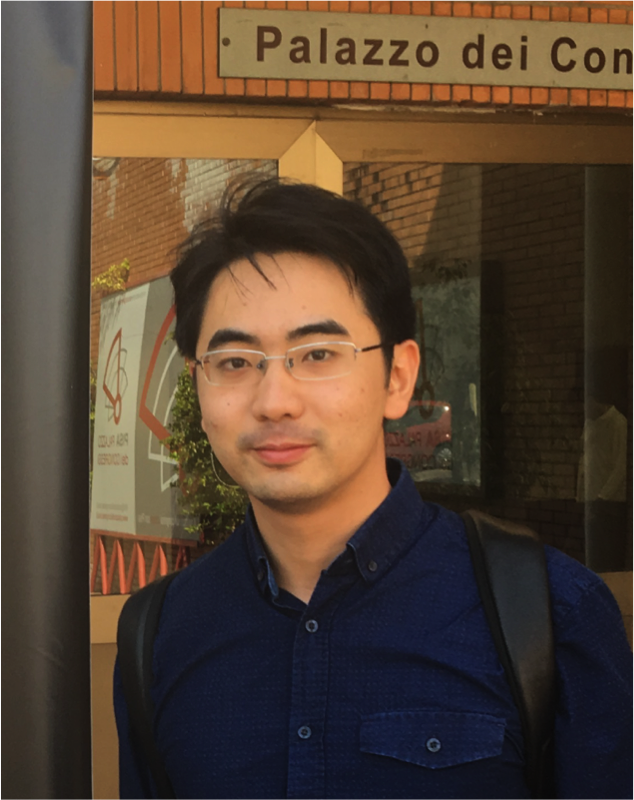}
Dr. Xiangnan He is a professor at the University of Science and Technology of China (USTC). He received his Ph.D. in Computer Science from the National University of Singapore (NUS). His research interests span information retrieval, data mining, and multi-media analytics. He has over 80 publications that appeared in several top conferences such as SIGIR, WWW, and MM, and journals including TKDE, TOIS, and TMM. His work has received the Best Paper Award Honorable Mention in WWW 2018 and ACM SIGIR 2016. He is in the editorial board of journals including Frontiers in Big Data, AI Open etc. Moreover, he has served as the PC chair of CCIS 2019 and SPC/PC member for several top conferences including SIGIR, WWW, KDD, MM, WSDM, ICML etc., and the regular reviewer for journals including TKDE, TOIS, TMM, etc.
\end{biography}

\begin{biography}{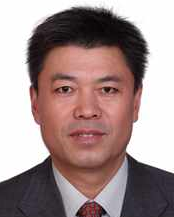}
Huamin Feng is now a professor at Beijing Electronic Science and Technology Institute. He received his Ph.D. dergree from National University of Singapore in 2005. His research interests include multimedia semantic analysis, recommend system, and web content analysis. He has published some academic papers on international conferences such as WWW, SIGIR, and MMM.
\end{biography}

\begin{biography}{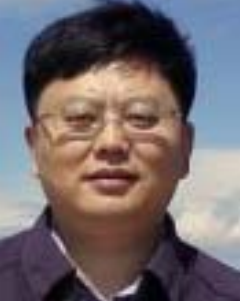}
Yongdong Zhang (Senior Member, IEEE) received the Ph.D. degree in electronic engineering from Tianjin University, Tianjin, China, in 2002. He is currently a Professor with the University of Science and Technology of China. He has authored more than 100 refereed journal and conference papers. His current research interests include multimedia content analysis and understanding, multimedia content security, video encoding, and streaming media technology. He was a recipient of the Best Paper Award in PCM2013, ICIMCS 2013, and ICME 2010; and the Best Paper Candidate in ICME 2011. He serves as an Editorial Board Member for Multimedia Systems journal and Neurocomputing.
\end{biography}

\end{document}